\documentclass[conference]{IEEEtran}
\IEEEoverridecommandlockouts
\usepackage{cite}
\usepackage{amsmath,amssymb,amsfonts}
\usepackage{algorithmic}
\usepackage{graphicx}
\usepackage{textcomp}
\usepackage{xcolor}
\def\BibTeX{{\rm B\kern-.05em{\sc i\kern-.025em b}\kern-.08em
T\kern-.1667em\lower.7ex\hbox{E}\kern-.125emX}}
\begin{document}

\title{Intelligent Stretch Reduction in Information-Centric Networking towards 5G-Tactile Internet realization\\
{\footnotesize}
\thanks{This work was supported by the National Research Foundation of Korea (NRF) grant funded by the Korean government (MSIT) (No. 2019R1A4A1023746, No. 2019R1F1A1060799) and Strengthening R$\&$D Capability Program of Sejong University.}
\thanks{*Corresponding authors: H. S. Kim (hyungkim@sejong.edu) and R. Ali (rashidali@sejong.ac.kr)}
}

\author{\IEEEauthorblockN{Hussain Ahmad, Muhammad Zubair Islam, Amir Haider, Rashid Ali$^{*}$, Hyung Seok Kim$^{*}$}
{School of Intelligent Mechatronics Engineering}\\
{Sejong University, Seoul (05006), Korea}\\
{\{hussainahmad, zubair\}@sju.ac.kr, \{amirhaider, rashidali\}@sejong.ac.kr, hyungkim@sejong.edu}}
\maketitle

\begin{abstract}
In recent years, 5G is widely used in parallel with IoT networks to enable massive data connectivity and exchange with ultra-reliable and low latency communication (URLLC) services. The internet requirements from user's perspective have shifted from simple human to human interactions to different communication paradigms and information-centric networking (ICN). ICN distributes the content among the users based on their trending requests. ICN is responsible not only for the routing and caching but also for naming the network's content. ICN considers several parameters such as cache-hit ratio, content diversity, content redundancy, and stretch to route the content. ICN enables name-based caching of the required content according to the user's request based on the router's interest table. The stretch shows the path covered while retrieving the content from producer to consumer. Reduction in path length also leads to a reduction in end-to-end latency and better data rate availability. ICN routers must have the minimum stretch to obtain a better system efficiency. Reinforcement learning (RL) is widely used in networks environment to increase agent efficiency to make decisions. In ICN, RL can aid to increase caching and stretch efficiency. This paper investigates a stretch reduction strategy for ICN routers by formulating the stretch reduction problem as a Markov decision process. The evaluation of the proposed stretch reduction strategy's accuracy is done by employing Q-Learning, an RL technique. The simulation results indicate that by using the optimal parameters for the proposed stretch reduction strategy.
\end{abstract}


\section{Introduction}
\begin{figure}[htbp]
\includegraphics[width=0.5\textwidth]{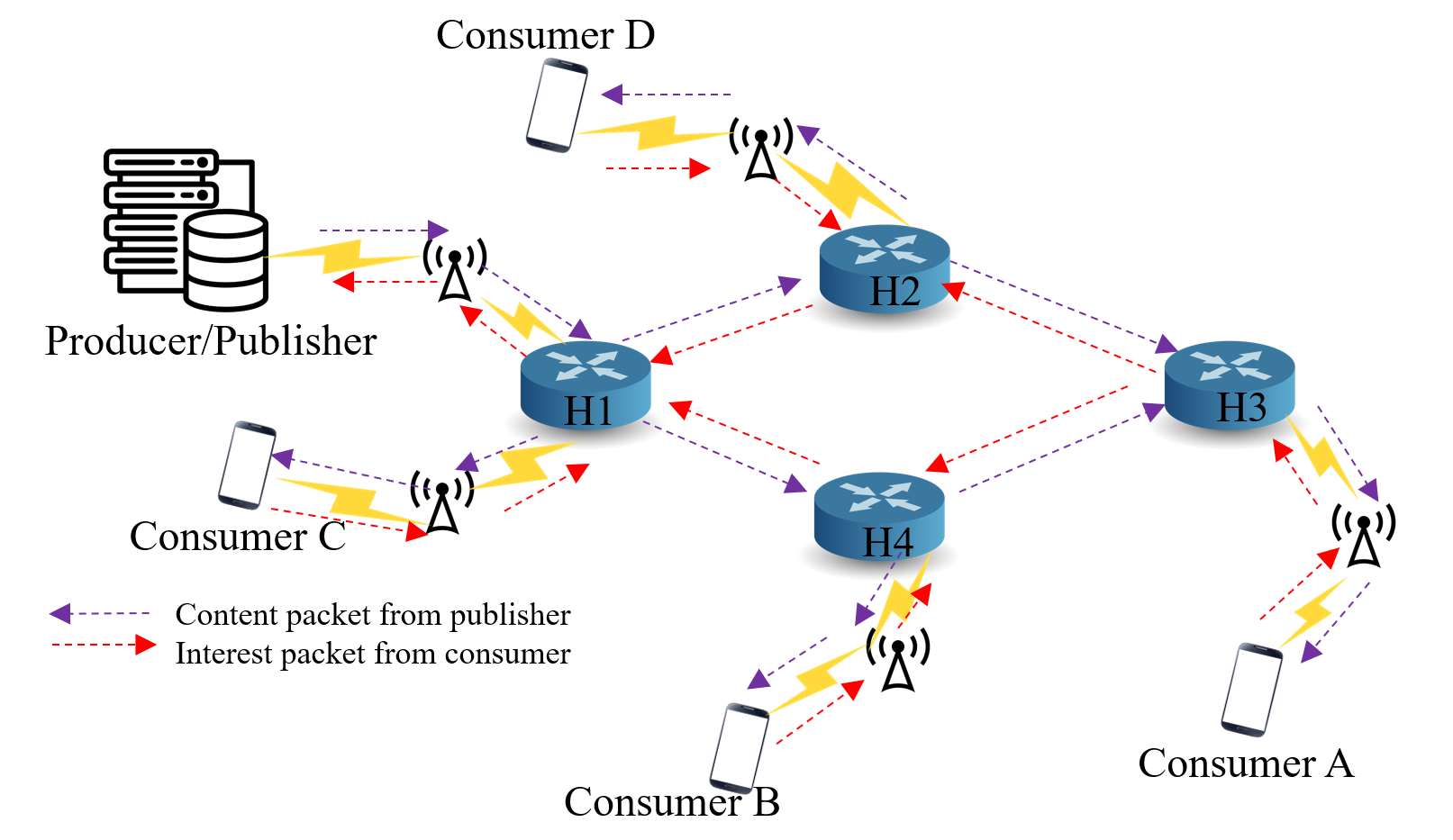}
\caption{Basic architecture of ICN system}
\label{fig:fig1}
\end{figure}
The fifth-generation (5G) network aims at providing services to enhance mobile broadband, massive machine-type communication, and ultra-reliable low latency communication (URLLC)\cite{b1}. 5G promises to provide a larger bandwidth for the users that need large bandwidth for their tasks. It also provides services for the delay and reliability of sensitive tasks. Similarly, 5G connects everything from everywhere, which is also named the Internet of Everything (IoE) \cite{b2}. In real-time applications, a massive number of devices are downloading data and uploading the data through various platforms. This leads to an increase in content and also the content diversity\cite{b3}. The content-type also varies in a way that it might be a video or sensor data. As ICN also introduces to handle the massive IoT user's data, it encouraged the researcher to introduce name-based networking called information-centric networking (ICN)\cite{b4}. It deals with the request of users from the name of the requested content. This paradigm is more concerned about "what" the data is rather than from "where" \cite{b5}. In ICN, the user generates an interest packet sent to ICN routers to check their cache memory or forward to the publisher to complete the content request and cache that content in cache routers. ICN routers have a forward interest base (FIB), content store (CS), and pending interest table (PIT) that help ICN for on-path and off-path caching. ICN allows data dissemination, ease of data access, and content level security\cite{b6}. Fig. \ref{fig:fig1} provides an overview of the ICN network. ICN provides more security, mobility, and in-network caching that makes the infrastructure neglect the location of the content. In-network caching minimizes the data traffic load on the core network while caching the data on routers\cite{b7}. ICN involves different cache strategies that cache the content in the routers available in the network topology. Numerous cache strategies like leave copy everywhere, leave copy down, maximum gain in-network caching and, ProbCache have their trade-offs between cache-hit-ratio (CHR), content diversity (CD), content redundancy (CR), and throughput. Other research groups are working to improve every single cache strategy to improve the network efficiency\cite{b8}.\par
\begin{figure*}[htbp]
\includegraphics[trim=0cm 1cm 0cm 2cm, clip=true,width=1 \textwidth]{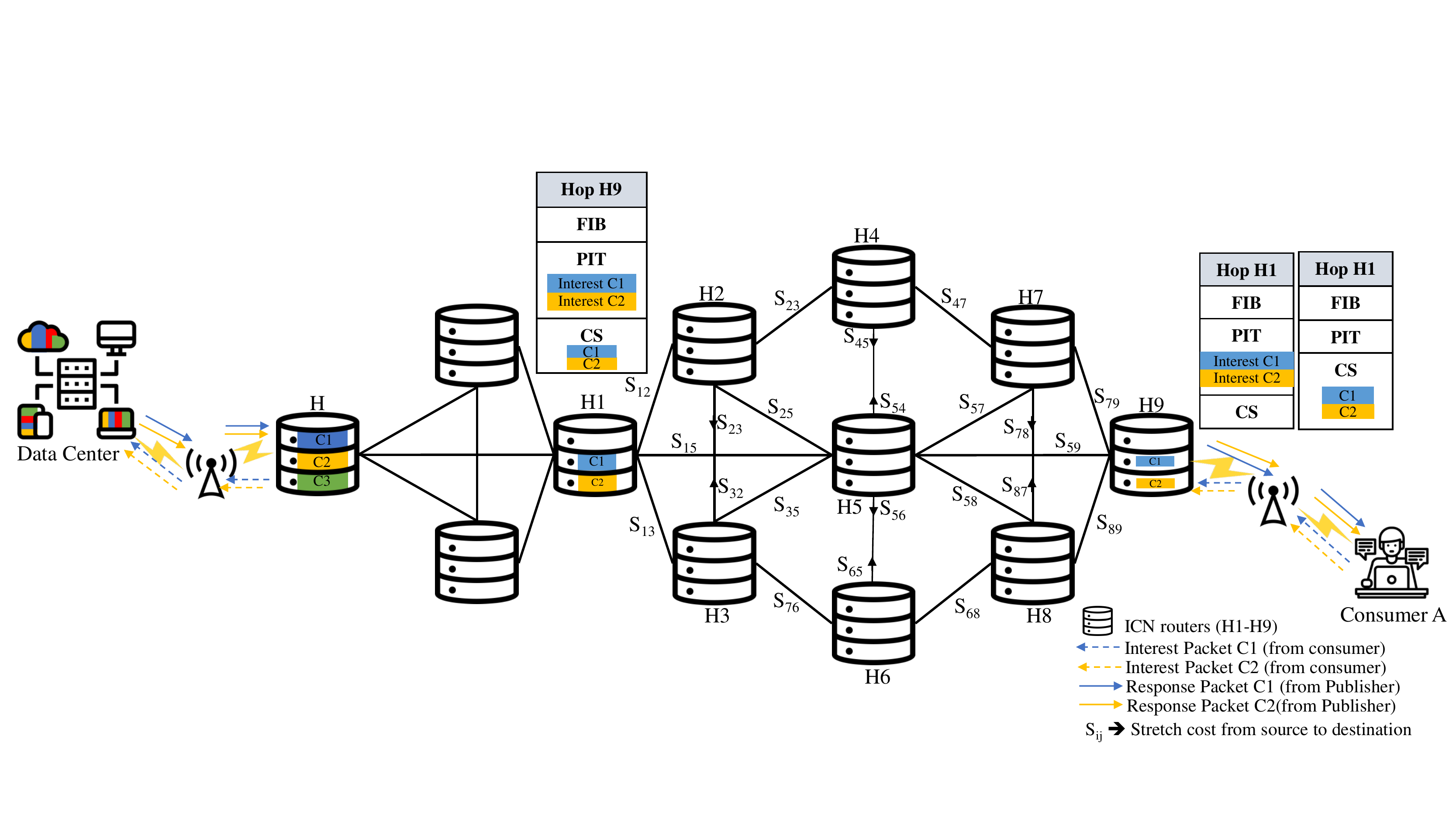}
\caption{RL-based stretch reduction framework}
\label{fig:fig2}
\end{figure*}
Several research groups have worked in the domain of ICN to increase the efficiency of the overall ICN infrastructure and discussed the parameters like CHR, CR, CD, and stretch \cite{b9,b10}. However, many authors have focused on CHR, latency, and throughput. The stretch parameter has been worked as a trade-off that if their infrastructure has attained better CHR, latency, and throughput, they neglected the parameter of a stretch to improve. Stretch is the distance traveled by the interest packet from the consumer (user, UE, or consumer) to the publisher (cloud, core network, or data server)\cite{b11}. Different cache strategies have been evaluated and have discussed their results in detail, while they work in the area of stretch reduction is not pursued in detail. Stretch also aids in increasing the overall network throughput in how many hopes/nodes the user request has traveled to retrieve the content. The lesser the number of hopes/nodes, the lesser the stretch which will aid latency rates and available data rates ate respective ICN router \cite{b12}.\par
Recently many authors have introduced cache strategies to improve the overall efficiency of ICN. In \cite{b13} authors have discussed and simulated the ICN based cache strategies like HPC, MPC, CCAC, WAVE, D-FGPC, CPCCS. An extensive survey and comparison were carried out. In \cite{b14}they have presented and evaluated an ICN based Edge Computing framework for 5G. they introduced ICN based Radio access network(RAN), D2D communication to avoid the minimal requests to be sent to core network considering the 5G infrastructure. They introduced the content prefetching strategy for the most popular content to be prefetched to the RAN. They presented and evaluated their proposed content prefetching idea and presented extensive work on 3 tier network topologies based on Edge computing framework, that is, D2D level, Edge Level, and RAN level. \cite{b15} presented ICN enabled forwarding devices to cache content at the BS and core network. Thus, this paper also focused on the increased number of CHR at edge devices and minimizing the requests to be forwarded to the core network. In \cite{b16} authors have proposed a novel cache strategy of Most interesting content caching (MICC). MICC aims to provide content dissemination and cache the transmitted content at a favorable location. MICC is evaluated with conventional parameters like CHR, stretch, Content eviction (CE), and CD. In \cite{b17} authors proposed a periodic caching strategy. They used between centrality router to cache the content closer to users. The content which is mostly being requested by the users is cached at the edge routers, and if the edge routers have not enough storage left, then the other popular content will be cached at the most central routers. This paper also aimed to improve CHR, but they also illustrated that they observed the reduction in stretch and content retrieval latency by implementing PCS.\par
Artificial Intelligence (AI) is an emerging tool to tackle the problem in the areas of caching and communication in multi-access edge computing-based 5G infrastructure \cite{b18,b19,b20}. In \cite{b18}, the authors discussed the transfer learning-based approach considering the dynamic behavior of the content and evaluated the time-varying content popularity to cache the contents in a small cell environment. Authors \cite{b19} considered a joint BS scenario for caching content using Q-learning. Q learning is also used for content replacement according to respective content popularity. in \cite{b20} authors have proposed an estimation based scheme for calculating the unknown popularity of the cached content. thus providing better QoS. However, the work done in ICN's domain using RL to reduce stretch is not much discussed and pursued.\par
In this paper, our research objective is to minimize the path length i.e. stretch between the user and the cached content when the content has been identified on which router it is placed. We introduced a Reinforcement Learning-based approach to reduce the stretch between the user and the content router. Therefore, our contributions are a step ahead by implementing AI in the ICN to learn based on exploration and exploitation to give better performance.\\
The remainder of the paper is organized as follows. Section II elaborates on our proposed RL-based ICN framework for stretch reduction. Section III consists of the performance evaluation, results, and discussion, and we concluded our paper in Section V with some future research directions.
\section{RL-based Framework}
We proposed our Reinforcement learning-based algorithm and the main goal of this is to reduce stretch reduction for the ICN framework. in this section, we first describe the RL; secondly we describe the Q-learning and Markov decision process (MDP) which we have used a Learning approach for our problem statement. Third, we describe our system model in detail while considering a case scenario to understand our approach better. 
\subsection{Reinforcement Learning}
In RL, a node learns to take actions and then maps possible outcomes or situations for the agent's actions. Generally, the node or device does not have the knowledge about what actions to perform.But the RL agent has to discover the best reward by exploring the possibilities. DRL has some preliminary elements like agent and environment. besides that, it has some policy which acts as a strategy. it characterizes the states of the respective environment to take actions relatively. the other elements include reward, value function, or environment model. The reward is the parameter that helps the policy to learn it better and calculate better policy. the value function is the sum of rewards encountered in the whole episode. The environment model shows how the environment behaves and what the conditions are.
\subsection{System Model}
Our RL-based stretch reduction algorithm mainly focuses on selecting the optimal path i.e., the path with fewer hopes or ICN routers. As RL learns from the experience, it will converge to a better policy after some time or iterations. Thus RL algorithms aim to reach to the optimal policy after learning from a number of experiences. Our scenario has the following entities: 
\begin{itemize}
\item User(consumer)
\item A content router or between centrality router-based ICN router having the contents that are of interest of a user
\item Other ICN enabled routers
\end{itemize}
Our proposed system model has been shown in Fig. \ref{fig:fig2}. We supposed that router H1 has the content according to the user connected with router H9. We have observed the mesh network of routers in our typology. all the routers H1-H9 are connected just like a mesh network. Consumer B is connected with router H9 wirelessly. In a mesh topology, almost every routing device is connected with each other, as illustrated in Fig. 2. 
$R_{ij}$ illustrates the stretch reward encountered by each step the agent has taken from its present or current say hope $H1$ to $H2$. 
\subsubsection{Markov Decision Process}
\begin{figure}[htbp]
\includegraphics[width=0.5\textwidth]{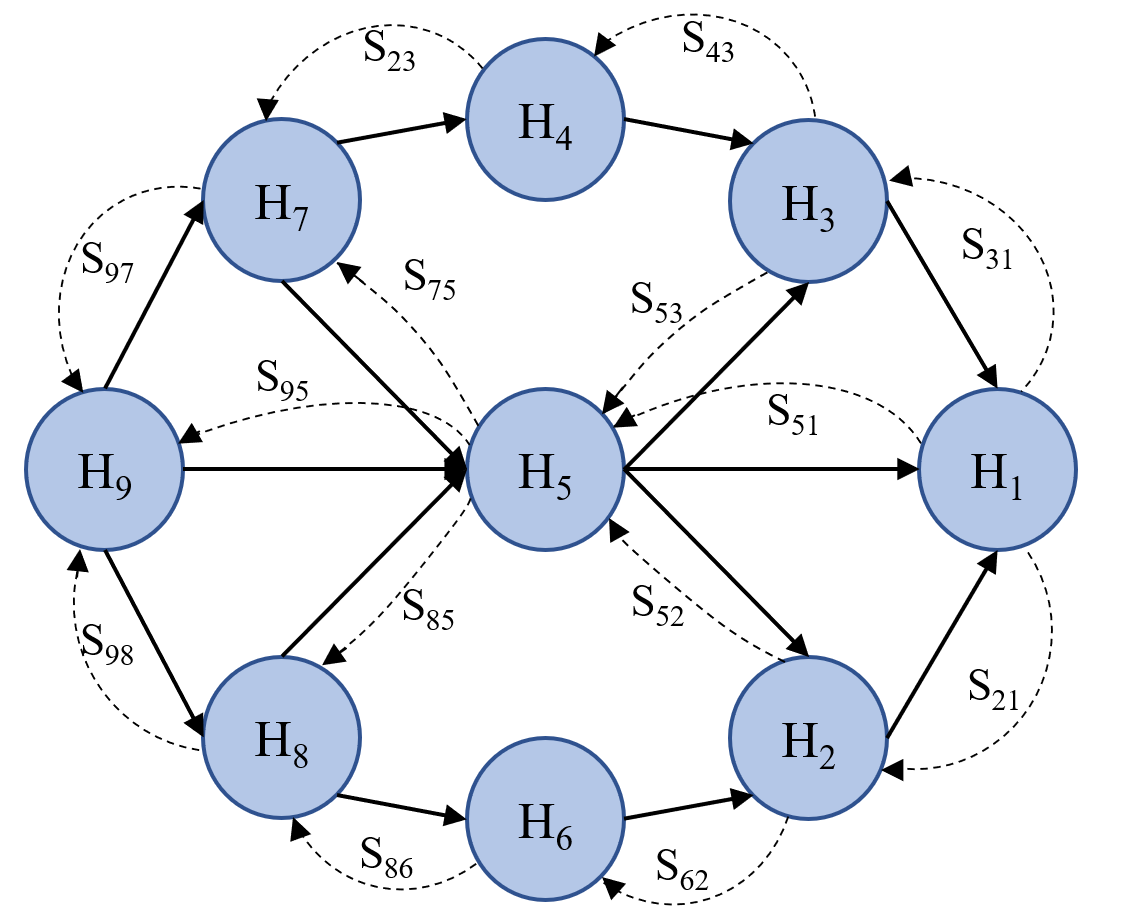}
\caption{MDP of proposed algorithm (an example of n=9 states)}
\label{fig:fig3}
\end{figure}

The process in which the agent observes the environment's output in parameters like a reward, next state, and then what action would be taken next is the Markov Decision Process (MDP). Fig. \ref{fig:fig3} shows that our proposed model also follows MDP.
\begin{equation}
P[H_{i+1} | H_i] = P[H_{i+1} | H_1, ….. , H_i]\label{eq1}
\end{equation}
the MDP for our router $H_i$at time t can be described as follows:
\begin{itemize}
\item The router $H_i$ senses its current state and obtains the current state 
\item Based on current state $H_i$ Router t i selects action $A_i$
\item Based on $A_i$ the environment makes a transition to new state $H_{i+1}$ and get a reward $R_i$
\item That reward $R_i$ is feedback to the router and the process continues to next states with the same process
\end{itemize}
\begin{figure*}[t]
\includegraphics[width=1 \textwidth]{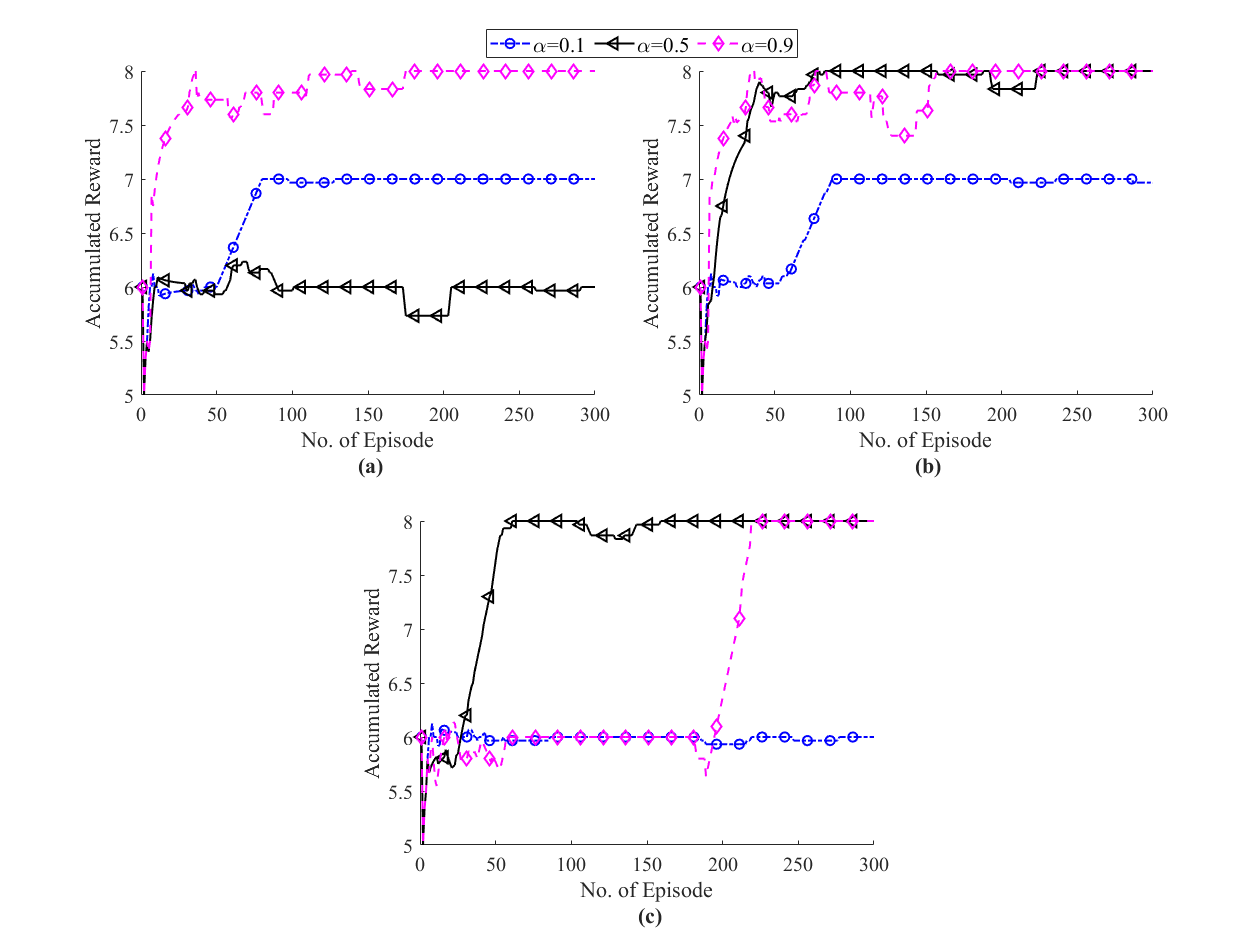}
\caption{Convergence to optimal stretch w.r.t (a) $\gamma$ = 0.1; (b) $\gamma$ = 0.5; (c) $\gamma$ = 0.9}
\label{fig:fig4}
\end{figure*}
\subsubsection{Q-Learning Model} 
Q-learning is an off-policy RL algorithm. It is used to solve the MDP. We modeled a stretch based problem as an MDP shown in Fig. 3 and proposed a Q-Learning based stretch reduction strategy.
First, we construct an MDP based on the stretch for mesh network topology. The MDP model can be defined by {H,A,R}
\begin{itemize}
\item $H_n$ is the set of possible states in our network topology from router $H_1$ to router next router.
\item $A_n$ is the set of actions that denotes which what are the required actions in our topology
\item $S_{ij}$ is the reward function that has the feedback of reward of what action has been taken by the agent
\end{itemize}
The goal is to minimize the stretch and reward shows the quality of step taken by the agent. thus we can evaluate the reward R in the send subsection.
to estimate the optimal value function, the optimal Q-Value is for router $H_i$ is defined as: 
\begin{equation}
Q_{i+1}(H_i,A_i)= (1-{\alpha})Q_i(H_i,A_i) + {\alpha} {{\triangle}Q_i (H_i,A_i)} \label{eq2}
\end{equation}
\begin{equation}
{\triangle}Q(H_i,A_i)= R_i+{\gamma} \max_{A'} Q_i(H',A') \label{eq3} 
\end{equation}

We have used the Q-Learning-based RL approach; therefore, it is mandatory to declare the environment with some states, action, policy, and rewards. Therefore, $H_n= [H_1,H_2,H_3,....,H_n]$ is state space in our infrastructure. Where, n=9 in our scenario. Similarly, the action space is defined as $A_n= [A_1,A_2,A_3,A_4,A_5]$.The action space defines the possible hop the content should follow to reach the consumer based on taking the policy. As discussed, the higher the reward minimum will be our stretch. thus the higher reward declares the minimum stretch in the whole topology. 
\begin{equation}
S_{ij} = 1/R_{ij}
\end{equation}
where $S$ is stretch. $RL_i$ is the number of routers or hops traveled by the interest packet consumer and publisher evaluated using the Q-learning.
\subsection{Simulation Environment}
We use MatLAB 2020a and consider nine states which are used as ICN routers, also illustrated in Fig. \ref{fig:fig2}. Our proposed scheme utilized the mesh network of routers as a network topology. The consumer is connected to H9, and the router H1 has the desired content in its cached memory, according to consumer H9. Our simulation results illustrate that the RL-based network will help the consumer retrieve the content from that path with less stretch that will help get a better data rate and lesser latency. Table \ref{tab1} illustrates the default parameter values of simulation environment.


\begin{table}[htbp]
\caption{Default Simulation Parameters}
\begin{center}
\begin{tabular}{|l|l|}
\hline
\textbf{Simulation parameters} & \textbf{\textit{Values}} \\
\hline
Number of producer & 1 \\
\hline
Number of consumer & 1 \\
\hline
Number of states (ICN Router) & 9 \\
\hline
Number of Actions (Routing paths) & 5 \\
\hline
Total Number of contents & 3 \\
\hline
Learning rate (${\alpha}$) & 0.1, 0.5, 0.9 \\
\hline
Discount factor (${\gamma}$) & 0.1, 0.5, 0.9 \\
\hline
Exploration/Exploitation (${\epsilon}$) & 0.5 \\
\hline
\end{tabular}
\label{tab1}
\end{center}
\end{table}

\section{Results and Discussion}
In this section, the performance of our proposed scheme is evaluated, which resembles the real network environment. Simulation results show that the RL-Based Stretch algorithm achieves the reduction and converges to optimal performance in the little amount of iterations.\par
To evaluate our RL-based Stretch reduction algorithm, we kept different parameters of $\alpha$, $\gamma$, $\epsilon$.
We observed RL algorithm's performance on stretch by varying the different values of $\alpha$, $\gamma$, $\epsilon$. Alpha is the learning rate in RL models.\par
We have kept the value of $\epsilon$ = 0.5 in all of our simulation results as it means that 50\% the model will explore and 50\% it will exploit. in Fig. \ref{fig:fig4}(a) it is seen that graph with $\alpha$ value after 100$^{th}$ iteration does not completely converges. In comparative to the $\alpha$ = 0.5 the results are better as compared to that of $\alpha$ = 0.1 as it convergence to better stretch. However, at $\alpha$ value 0.9, the maximum convergence has been observed that results in the better and optimal stretch in this case of $\gamma$=0.1. But the model has attained the convergence at approximately 180th iteration.\par
In Fig. \ref{fig:fig4}(b) two out of one results have converged to optimal stretch. The parameters are all the same as discussed above but with the change of $\gamma$ value. We have changed the $\gamma$ value to 0.5 and observe the trend. It can be observed clearly that that at $\alpha$ value 0.1, the model converged to better stretch, and at $\alpha$ = 0.5, the model completely converged to optimal stretch at 267$^{th}$ iteration. While the $\alpha$ value of 0.9 converged fastly at 165$^{th}$ iteration as compared to that of $\alpha$ value 0.5. \par
In Fig. \ref{fig:fig4}(c) the better performance has been achieved as the $\alpha$ 0.5 and 0.9 converged to optimal stretch. However, the $\alpha$ value has converged lately as compared to that to Fig. \ref{fig:fig4}(b).\par
The parameter $\gamma$ at 0.5 and $\alpha$ at 0.5 converges fastly towards the optimal stretch. Therefore, with careful analysis, it has been illustrated that our model drives the agent to provide minimum stretch after some iterations. 
\section{Conclusion and Future Discussion}
In this paper, our research objective is to minimize the path length i.e., stretch between the user and the cached content when the content has been identified on the router. We have introduced Q-Learning based approach that is the technique of Reinforcement Learning to reduce the stretch between the user and content router. Therefore, we have implemented AI in the ICN to learn based on exploration and exploitation to give better performance. Moreover, we construct an MDP framework for the stretch reduction problem for ICN and Q learning approach to solve the MDP that reduces stretch for ICN router when the content is retrieved from publisher to user. The proposed algorithm uses RL based strategy to converge to the optimal path for the ICN framework. The convergence for different parameters is illustrated and simulated, proving that our proposed scheme converged to the optimal path with a minimum number of iterations. 

\vspace{12pt}

\end{document}